\documentclass{spie}
\usepackage{geometry}                % See geometry.pdf to learn the layout options. There are lots.
\usepackage{graphicx}
\usepackage{amssymb}
\usepackage{epstopdf}
\title{Stellar intensity interferometry: \\Experimental steps toward long-baseline observations}
\author{Stephan LeBohec\supit{a}\footnote{ \ \ lebohec@physics.utah.edu},  Ben Adams\supit{a}, Isobel Bond\supit{b}, Stella Bradbury\supit{b}, Dainis Dravins\supit{c},\\ Hannes Jensen\supit{c}, David B. Kieda\supit{a}, Derrick Kress\supit{a}, Edward Munford\supit{a},\\ Paul D. Nu\~nez\supit{a},   Ryan Price\supit{a}, Erez Ribak\supit{d},  Joachim Rose\supit{b},  Harold Simpson\supit{a}, Jeremy Smith\supit{a}
\skiplinehalf
\supit{a} Department of Physics \& Astronomy, University of Utah, 115 South 1400 East, Salt Lake City, UT 84112- 0830, USA\\
\supit{b} School of Physics and Astronomy, E.C. Stoner Building, The University of Leeds, Leeds, LS2 9JT, UK\\
\supit{c} Lund Observatory, Box 43, SE-22100 Lund, Sweeden.\\
\supit{d} Physics Department, Technion, Haifa 32000, Israel\\
}
\begin{document}
\maketitle

\abstract{Experiments are in progress to prepare for intensity interferometry with arrays of air Cherenkov telescopes. At the Bonneville Seabase site, near Salt Lake City, a testbed observatory has been set up with two 3-m air Cherenkov telescopes on a 23-m baseline. Cameras are being constructed, with control electronics for either off- or online analysis of the data. At the Lund Observatory (Sweden), in Technion (Israel) and at the University of Utah (USA), laboratory intensity interferometers simulating stellar observations have been set up and experiments are in progress, using various analog and digital correlators, reaching 1.4 ns time resolution, to analyze signals from pairs of laboratory telescopes.}

\section{Introduction}
The recent years have seen an increase of interest for the possibility of reviving Stellar Intensity Interferometry (SII), which was pioneered by Robert Hanbury Brown and Richard Twiss who established the technique in 1956 \cite{rhbandrt1956}.  They then exploited\cite{rhb1974a} it with their colleagues  with the Narrabri Stellar Intensity Interferometer (NSII \cite{rhb1974b}) operated until 1972, after which the technique was abandoned. The present renewed interest draws from a few key ideas. Large baselines ($100\,m$ and more) in the visible band are still challenging for amplitude (Michelson) interferometry while the SII technique is relatively unaffected by poor seeing conditions and large baselines. Several large scale projects including large arrays of large light collectors are currently under development. Technological developments in photo-detection and signal processing since the time of the NSII make it possible to process the high bandwidth signals from a large number of telescopes. 

The recent interest for SII induced the formation of a dedicated working group within IAU commission 54 \cite{olbin,spie2008} which had a first meeting in January 2009\cite{groupmeeting} during which planned imaging atmospheric Cherenkov telescope (IACT) arrays used for very high energy ($E>100\,GeV$) gamma ray astronomy were identified as the most favorable settings for the deployment of a major Stellar Intensity Interferometer. The working group subsequently contributed to the 2010 Astronomy and Astrophysics Decadal Survey Astro2010 in the USA with a white paper\cite{wp2010} and a response to a request for information\cite{rfi2010} identifying the science potential of a modern version of SII with large telescope arrays as well as technical challenges and options for a successful deployment.  This program is now underway as SII gets increasingly integrated in IACT projects in the form of task force groups within the CTA\cite{cta} and  AGIS\cite{agis} collaborations which are planning construction to start in 2013. The recently approved upgrade of the VERITAS gamma ray observatory \cite{veritas} includes SII in its program and will serve as a test bench for more ambitious SII projects. Efforts are going along four main directions  about each of which a report can be found in these proceedings. The science potentials of a modern, large scale SII with imaging capabilities are being better identified and characterized\cite{dainis} with lists of interesting targets. Imaging possibilities offered by SII are implemented and evaluated\cite{paul} using increasingly realistically simulated data. The optimal geometry of telescope arrays to be used as SII is investigated in order to identify and evaluate possible options and compromises that would further improve SII capabilities without being detrimental to the Very High Energy (VHE -- $E>100\,GeV$ ) observing program\cite{hannes}. Finally, the present paper is a report on developments concerning the design and testing of various prototypes for the optics, front end electronics and correlation electronics. 

This paper is organized as follows. Section \ref{sii} gives a brief outline of the principle of intensity interferometry with its capabilities and requirements using IACT arrays. Section \ref{sbt} presents the telescopes in StarBase which are to be used as a realistic environment testing bed with easy access.  Section \ref{camerasec} presents a prototype of secondary optics making SII compatible with VHE instrumentation.  Section \ref{electronics} presents the slow control and front end electronics prototypes under test. Individual photodetector signals are brought to the central recording and correlation station via analog optical fibers. Section \ref{correl} presents various correlator options being investigated. This is the critical and most challenging part of an SII as the instantaneous signal to noise ratios are extremely small. Finally section \ref{outlook} anticipates on the future developments toward the effective deployment and operation of an array of SII receivers. 

\section {Intensity Interferometry}
\label{sii}
Intensity Interferometry relies on the fact that the beating of Fourier components of light results
in correlated star light intensity fluctuations in different telescopes. The degree of correlation between the intensity fluctuation $\delta i_1$ and $\delta i_2$ recorded by two telescopes provides a measurement of the squared mutual coherence, $|\gamma(d)|^2$, of the light at the two telescopes separated by a distance $d$,  which is the square of the normalized magnitude of the Fourier transform of the image\cite{labeyrie}. The degree of correlation is obtained by averaging over time the product of the intensity fluctuations (see Equation \ref{correleq}) and the measurement relies on a second order effect implying severe sensitivity limitations which can only be alleviated by using telescopes with very large effective light collecting areas.
\begin{equation}
|\gamma(d)|^2= {{<\delta i_1 \cdot \delta i_2 >}\over{<i_1><i_2>}}
\label{correleq}
\end{equation}
 However, the intensity fluctuations are of relatively low frequency set by the exploited coherence time and the requirements on optical and mechanical tolerance are much relaxed in comparison to amplitude interferometry. The signal to noise ratio in a single pair of telescopes is given by Equation \ref{snratio} where $A$ is the light collection area of each telescope, $\alpha$ is the quantum efficiency, $n$ is the spectral density ($m^{-2} s^{-1} Hz^{-1}$), $\Delta f$ is the intensity fluctuation signal bandwidth and $T$ is the duration of observation. 
\begin{equation}
(S/N)_{RMS}=A \cdot \alpha \cdot n(\lambda) |\gamma(d)|^2 \sqrt{\Delta f \cdot T /2} = \sqrt{<i_1><i_2>}\cdot|\gamma(d)|^2\cdot \sqrt{\Delta f \cdot T /2}
\label{snratio}
\end{equation}
This equation does not take into account effects such as photo-detector excess noise, night sky background light or partial resolution by individual telescopes\cite{apj2006}. It can however be used to estimate the sensitivity of a realistic pair of telescopes. With $A=100\,m^2$, $\alpha=30\%$ and $\Delta f=100MHz$, a single pair of telescopes used for five hours would provide a five statistical standard deviation detection and measurement of degrees of mutual coherence $|\gamma|^2=0.3$ and  $|\gamma|^2=0.03$ for stars of visual magnitude 4.8 and 2.4 respectively \cite{rfi2010}. Crude light collectors composing IACT arrays used for VHE gamma ray astronomy observations are perfectly suited to be used also as SII receivers. VHE gamma ray observations are restricted to low moon light nights during which SII observations could be made through narrow optical passbands. The recent successes in the field of VHE gamma ray astronomy resulted in the ongoing design of much larger arrays counting up to one hundred telescopes which would provide close to five thousand simultaneous baselines \cite{hannes} ranging from $\sim 50\,m$ to more than $1000\,m$. Such a set of baselines with large telescopes suggests observations of visual magnitude 9 stars with an angular resolution of $50\mu as$ for $\lambda=400nm$ would be possible. The magnitude limitation is coming from both the practicality of the required observations times ($T<50\,h$) and from the night sky background signal contamination\cite{apj2006}. The rich coverage of the interferometric plane allows model independent imaging\cite{paul} by means of various phase recovery techniques\cite{holmes1,holmes2}.    

In order to make it possible to use IACT arrays for SII observations, dedicated instrumentation to measure, communicate, record and correlate the intensity fluctuations at individual telescopes is necessary. The subsequent sections of this paper are dedicated to the ongoing\cite{htra2008} design, construction of testing of prototypes aiming at providing SII capability to the future IACT arrays for which start of constructions is anticipated in the coming three years.

\section{The Star Base Telescopes}
\label{sbt}
\begin{figure}
\begin{center}
\rotatebox{0}{\includegraphics[scale=0.55]{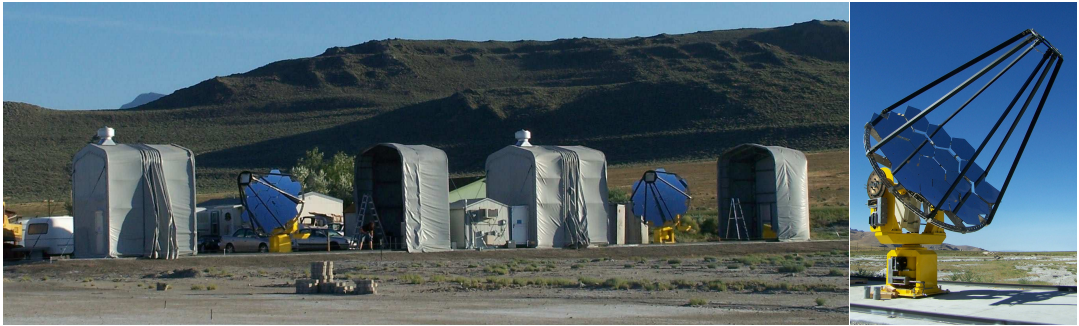}}
\end{center}
\caption{\label{starbase} The StarBase $3\,m$ telescopes are protected by buildings which can be rolled open for observation (left) The control room is located in a smaller building located between the two telescopes. The picture on the right shows a close up view of one telescope before the camera was mounted.  }
\end{figure}

As a first test toward implementing SII with IACT arrays, pairs of 12\,m telescopes in the VERITAS array at the Fred Lawrence Whipple Observatory in Arizona were interconnected through digital correlators \cite{dainisspie}.  These tests were made during parts of nights otherwise shared with VHE observations with a very temporary setup and established the need for a dedicated test bench on which various options of secondary optics and electronics could be evaluated in a realistic environment. In order to satisfy this requirement, the two StarBase\cite{starbaseweb} telescopes were deployed on the site of the  Bonneville Seabase diving resort \cite{seabase} in Grantsville, Utah, 40 miles west from Salt Lake City. The two telescopes (Figure \ref{starbase}) are on a $23\,m$ East-West base line. The telescopes had earlier been used in the Telescope Array experiment\cite{telescopearray} operated until 1998 on the Dugway proving range. Each telescope is a $3\,m$, $f/1$ Davies-Cotton\cite{daviescotton} light collector composed of 19 hexagonal mirror facets $\sim 60\,cm$ across. This design is typically used for IACT and secondary optics tested on the StarBase telescopes could be used directly on the VERITAS telescopes for larger scale tests. The telescope mounts are alt-azimuthal with the motion around both axes controlled by tangential screws and absolute encoders with a few arc-seconds resolution. The tracking model parameters are being optimized but the absolute pointing accuracy is better than four arc-minutes and can be compensated by online corrections. This should be compared to the focal plane optical point spread function (PSF) full width at half maximum (FWHM) which is on the order of $0.1^\circ$. The PSF is dominated by spherical aberration of individual mirror facets. This is untypical of large Davies-Cotton Cherenkov telescopes for which the PSF FWHM is typically around $0.05^\circ$. This difference is due to the facets of the StarBase telescope being much larger in proportion to the telescope diameter than in usual IACTs. For example, the VERITAS telescopes are $12\,m$ in diameter $f/1$ light collectors made of 350 mirror facets $\sim 60\,cm$ across. Interestingly, this lower angular performance of the StarBase light collectors make them ideal test beds for a larger scale implementation of SII as the PSF linear extent measured in the focal plane is very comparable to that in large IACTs such as in VERITAS and the aperture ratio is the same.    

\section{Stellar Intensity Interferometry Cameras}
\label{camerasec}
The cameras designed for the StarBase telescopes must satisfy two objectives. They must perform as SII cameras providing two channels for zero baseline measurement. They must also demonstrate their compatibility with the VHE Cherenkov camera occupying the focal plane of IACTs. This is achieved by using a large enough mirror making a $45^\circ$ angle with the telescope optical axis so all the secondary optics is mounted in a plane parallel to the focal plane. The idea is that such a SII camera could simply be mounted in front of the VHE camera or, even better, integrated in the shutter protecting the VHE camera. 

The SII camera must collect the light, selecting a narrow optical passband and concentrate it on one or two photodetectors if the zero baseline correlation is to be measured. Interest for narrow optical passbands is two fold. As already seen above, the sensitivity of SII does not depend on optical passband width. For very bright stars, selecting a narrow optical passband preserves the sensitivity while minimizing problems associated with measuring a high photon rate with a high gain photomultiplier tube. For faint stars there is no such need for a narrow passband. However the capability of working with narrow optical passbands will allow measurements through multiple independent optical passbands, thereby improving sensitivity for observations where wavelength dependence is not critical.  For example, replacing the beam splitter used to illuminate the two channels with a dichroic mirror would allow the measurement in two different optical passbands simultaneously. 

\begin{figure}
\begin{center}
\rotatebox{0}{\includegraphics[scale=0.35]{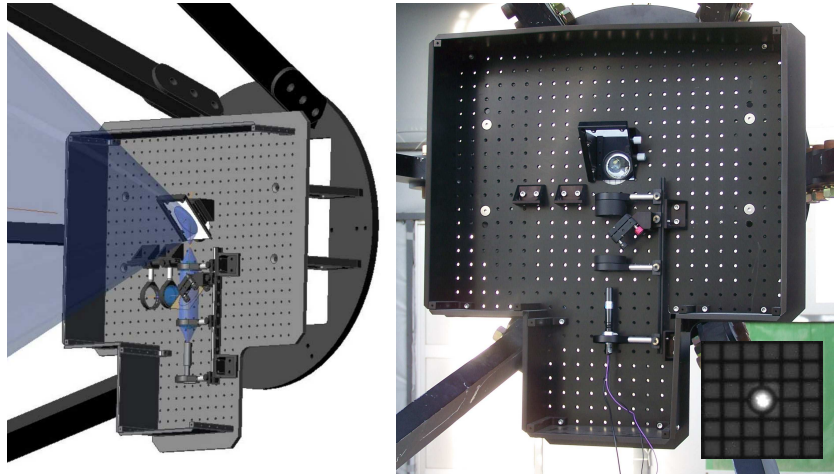}}
\end{center}
\caption{\label{camera} All the optics of the SII camera is mounted on a plate, just above the focal plane of the telescope as shown on the AutoCAD drawing on the left. The light is reflected by a flat mirror at a $45^\circ$ angle from the optical axis. It is then picked up by a collimating lens on which is mounted an interferometric filter. A beam splitter allows the light to be sent to two independent channels. The light is concentrated onto the photo-detector by an f/1.2 $60\,mm$ lens. The electronics can be mounted next to the photo-detectors and on the back of the camera. The actual camera is shown on the right with only one channel in place and no electronics yet.  The picture in the lower right corner is an image of the PSF taken while tracking Capella, The grid shows a 1\,cm spacing in the focal plane, which corresponds to $0.19^\circ$.}
\end{figure}

Working with narrow optical passband, say $\Delta \lambda \approx 10\,nm$ typically requires the light to be collimated to within $\sim 5^\circ$.  The extent of the PSF measured in the focal plane being $\sim 10\,mm$, this implies using a collimating lens of focal length longer than $\sim57\,mm$ and since the aperture ratio of the telescopes is unity, the secondary optics need to be of diameter larger than $57\,mm$. This simple estimate does not fully prevent vignetting of some of the point sources in the focal plane, $5\,mm$ from the optical axis so the minimal optics diameter should in fact be set to $67\,mm$. However, and in order to save some of the optics cost, we decided to work with $50\,mm$ optics, setting the focal length of the collimating lens to $37\,mm$ by combining a $100\,mm$ with a $60\,mm$ focal length lens. This results in a collimation within less than $8^\circ$. Figure \ref{camera} shows an autoCAD drawing of the camera and an actual camera mounted on one of the StarBase telescopes.

\section{Front end and slow control electronics}
\label{electronics} 
The camera electronics consist of two parts. The slow control electronics provide power to the front end, digitize the anode current to monitor the DC light intensity $<i_n>$, provide high voltage to the photo-detector and can be used to program parameters of the front end electronics. The front end electronics convert high frequency intensity fluctuations $\delta i_n$ down to the single photon level to analog light pulses which can be transported by optical fiber with minimal bandwidth loss over great distances. The optical fiber signals are then converted back to electrical signals which may be correlated at the central control building. 

\subsection{Slow control electronics}
The slow control system is responsible for providing high voltage power to the camera photomultiplier tubes (PMT) as well as capturing the average anode current $<i_n>$. A central microprocessor on the circuit board records the anode current that is optically isolated from the separate capture circuit through a built in ADC. The optical isolation is intended to avoid the injection of correlated noise in the front end electronics. The PMT is powered by an onboard programmable high voltage power supply that is controlled by the microprocessor through a DAC.  The circuit also records the ambient temperature and measures the voltage of its power source.  In order to adjust compensation for the temperature sensitivity of the gain of the front end electronics, the slow control system writes to a programmable resistor chip on the front end electronics through a two wire Inter-Integrated Circuit ($I^2C$) protocol.  The anode current, high voltage output, temperature, and power supply voltage data are relayed to a computer using the RS-232 protocol through serial to fiber-optic converters. The graphical software for the slow control board is written in wxPython and can control up to eight boards at a time. The software also records the anode current levels and writes them to a file so their values can be used in the data analysis to obtain measurements of $|\gamma(d)|^2$ by means of Equation \ref{correleq}. In the event that the PMT is exposed to  excessive   light levels, hardcoded circuit protection shuts off power to the high voltage supply. The slow control system conditions and provides power to the front end electronics. Each channel (two per telescope) is powered by local rechargeable batteries and is only optically connected to the outside of the focal plane instrumentation. 

\subsection{Front end electronics}
Future IACT arrays consisting of several tens of individual reflectors could in principle allow up to a few thousand different baselines. Detectors may be separated by as much as a kilometer, over which distance coherence must be maintained if one is to combine pulses of $3-5\,ns$ FWHM at a centrally located correlator. From Equation \ref{snratio}, we see that the sensitivity for SII observations is proportional to to the square root of the signal bandwidth.  Clearly signal transmission via co-axial cable which adds 50\% dispersion to the pulse FWHM and perhaps 25\% amplitude attenuation over $100\,m$ is not suitable\cite{htra2008}. 

The concept of a VCSEL (Vertical Cavity Surface Emitting Laser) based system to transmit undigitized photomultiplier pulses over optical fibre at high bandwidth with minimal attenuation for use in IACTs was proposed by J. Rose\cite{rose2000}. Optical fibres have the additional advantage of immunity to cross-talk and to electromagnetic interference and avoid the difficulty of maintaining a common ground and protection for the receiving electronics against  (not uncommon) lightning strikes across the array. They are also relatively lightweight. A 12 channel transmitter of the type tried in the Whipple $10\,m$ telescope weighed 650\,g, mainly due to its heatsink, and the weight per channel of the 30 core graded index glass fibre cable used was about a tenth of that of RG59 copper cable \cite{bond2001}. Thus the short-term installation, for interferometric observations in several optical passbands, of a plate carrying secondary optics, photomultipliers and signal transmission electronics in front of an existing IACT camera is conceivable. 

The single channel optical fibre signal transmission system being developed in Leeds for deployment between SII instrumentation prototypes is essentially a revision of that described in detail by R.J. White \cite{white2008}, which was found to give a linear response to within 12\% over a dynamic range of 61 dB. In the transmitter, an operational amplifier is used to develop a signal current proportional to the input signal voltage which is then AC coupled to the $850\,nm$ Zarlink ZL60052 VCSEL anode. The bias current supplied to the VCSEL, by a MAX3740A laser-driver chip, can be controlled by the slow control system over an I$^2$C interface via a DS1859 resistor chip. The DS1859 also monitors temperature and contains a look-up table which can be pre-programmed to maintain a uniform gain by adjusting the bias current in response to the self-heating of the VCSEL. Furthermore, the VCSELs used have been carefully selected from a batch tested for stability in order to avoid the sporadic gain changes observed in the Whipple $10\,m$ telescope system and attributed to laser mode hopping. 
The signal is communicated via a $62.5\,\mu m$ core, multi-mode optical fibre coupled to a PIN photodiode based receiver. Both VCSEL and photodiode are encapsulated in E2000 connectors for laser safety. 

An RC network and amplifier has been added to the transmitter board to convert the anode DC current to a voltage made available on an RJ45 connector for the slow control system used to provide power, monitoring and control signals. As well as providing the normalization for the measurement of the degree of coherence $|\gamma(d)|^2$, this allows one to record changes in the apparent brightness of a target star due to atmospheric effects and to safely reduce or cut the high voltage supplied to the photomultiplier should stray light flood its field of view. Further additional functionality includes the ability to inject test pulses into the laser-driver to test the link in daylight. The OPA695 amplifier used by R.J. White et al.\cite{white2008} has been superseded by Analog Devices' AD8000, one of a new generation of amplifiers not previously available, which has a comparable slew rate but a quoted input voltage noise of just $1.6\,nV/\sqrt{Hz}$. Initial laboratory tests show that a photomultiplier pulse rise time of the order of $2 ns$ can be well reproduced.

\section{Correlators}
\label{correl}
One of the most critical elements of an SII is the correlator which provides the averaged product of the intensity fluctuations $<\delta i_1 \cdot \delta i_2>$ to be normalized by the average intensities $<i_1>$ and $<i_2>$ provided by the slow control system. Several approaches are investigated in different groups. Different systems are currently being tested in the laboratory at the Lund Observatory, at the University of Utah and in Technion. These laboratory tests also include experimentation with different types of detectors from high quantum efficiency photo-multiplier tubes to single-photon-counting avalanche photodiodes. In these laboratory experiments, an artificial star is provided by an illuminated pinhole at some distance from miniature telescopes whose separation can be varied by moving them on an optical bench perpendicular to the observing direction. Photodetectors and correlators will be further tested on the StarBase telescopes starting summer 2010. 

\subsection{Analog correlator}
\begin{figure}
\begin{center}
\rotatebox{0}{\includegraphics[scale=0.75]{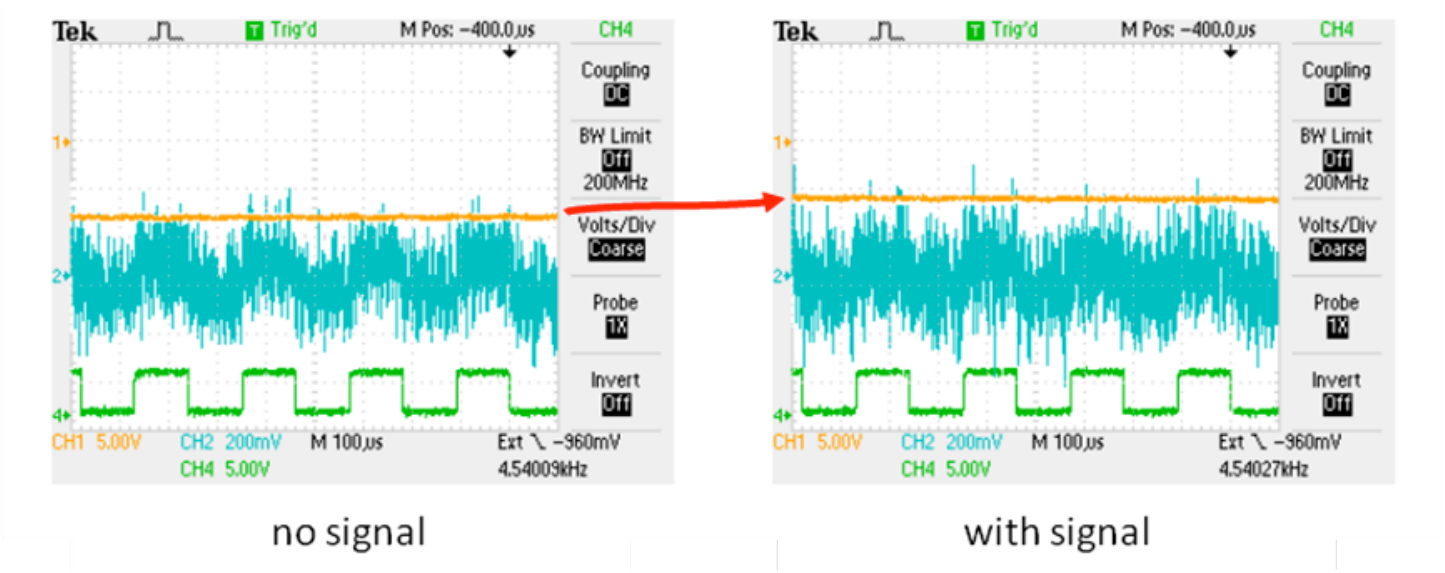}}
\end{center}
\caption{\label{technion} (Left) A modulator at the output of the detector (bottom curve) modulates the signal (middle), and the result is demodulated after the correlator (top). (Right) In the presence of signal, the noise increases, as is obvious from the middle and top curves. This additional noise is actually our correlation output. }
\end{figure}

At the Technion, a laboratory experiment is set up for testing the applicability of intensity interferometry to space, where large collectors and long base-lines are relatively easier to deploy. Communications between collectors (perhaps on different spacecrafts) might be the bottleneck. In the laboratory, a barely resolvable blue LED is monitored by two Fresnel lenses and photomultiplers. Each photomultiplier current signal is converted to a voltage, and the two voltages are correlated by an analog device (Figure \ref{technion}). This is a communications RF gain and phase detector, able to reach $2.5\,GHz$. This device serves as a correlator able to find the phase difference between the two inputs. It is operated above $100\,MHz$ because of interference from local radio stations, and below $1\,GHz$ because of detector limitations. The phase signal is found assuming we know approximately the frequency of the inputs, which is of course random. However the frequency is not so important as long as we stay in the zero-phase-difference regime.

For bright objects as encountered in Intensity Interferometry, the Poisson signal is overwhelming, and only deviations from it signify a signal. For example, a 20-bit number of photons will yield a 10-bit Poisson noise. Obviously deviations from Poisson noise on the scale of one or two bits are insufficient, so an even higher signal might be required. This is still difficult to handle with digital electronics, which is why analog processing is the initial choice. 

At the same time, for the space application, we need to transmit the signals to a central correlator. Here we opted for 8-bit digitizers running at $1\,GHz$, followed by correlation in  Field programmable Gate Array (FPGA). This version is now being tested and integrated.

\subsection{Digital on-line correlator}
A $100\,MHz$ signal bandwidth correlator was designed and constructed at the University of Utah. The two input correlator digitizes the signals into 12\,bits at a rate of $200\,MHz$. The digitized signals are collected by an FPGA programmed to provide a $5\,ns$ resolution time delay. In order to obtain a finer control of the relative timing, the FPGA also programs analog delays in st  of $0.6\,ns$ up stream for the digitization. After the digital delay, the samples are multiplied and summed up in an accumulation register to obtain the correlation. In order to minimize the effects of offsets and slow drifts, the correlator makes use of a double phase switching. The FPGA controls a programmable periodic phase modulation on both inputs, prior to digitization.  The multiplied samples are in fact added or subtracted from the accumulations register in a way that depends on the mutual states of the phase switches do achieve a demodulation of the correlation signal. This system is now fully operational. 

\subsection{Photon stream correlator} 

The Narrabri intensity interferometer is often seen as the first
experiment in quantum optics, and its subsequent theoretical
understanding, together with the development of the laser, led to a number
of somewhat analogous applications of light scattering against laboratory
specimens \cite{becker2005} 
It was realized that high-speed photon correlation measurements were
required and electronics initially developed in military laboratories were
eventually commercialized, first by Malvern Instruments Ltd. in the U.K.\cite{pike1979}.

At the Lund Observatory, such digital correlators were acquired and used to
pursue experiments for high-speed optical astrophysics, including studies
of atmospheric scintillation at the observatory on La Palma \cite{dravins1997}.
 While the first correlators were
impressively voluminous rack-mounted units, their electronics have since
been miniaturized and current units are very small and easily
transportable items, built around FPGAs,
accepting many input channels, running at sampling frequencies up to $700\,MHz$, 
and with easy USB data transfer to a host computer.  It is believed
that their electronic performance is now adequate for full intensity
interferometry experiments.  These have already been used at different
observatories, both in searches for high-speed astrophysical phenomena,
when connected to the OPTIMA photometric instrument of the
Max-Planck-Institute of Extraterrestrial Physics \cite{kanbach2008}, and also for the first
full-scale intensity interferometry experiments with Cherenkov telescopes
of the VERITAS array in Arizona \cite{dainisspie}.

At present, different correlators (made
by the company Correlator.com) are used, with the highest time resolution
reaching $1.4\,ns$.  The correlators can handle continuous data flows of more than  $100\,MHz$ 
per channel without any deadtimes.  Their computer output
contains the cross correlation function between the two telescopes (as
well as autocorrelation functions for each of them), made up of about a
thousand points.  For small delays (where most of the intensity
interferometry signal resides), the sampling of the correlation functions
is made with the smallest timesteps, which increase in a geometric
progression to large values to reveal the full function up to long delays
of seconds and even minutes.  Individual photon events are normally not
saved, although that is possible for moderate count rates below about $1\,MHz$.  
The advantage with such real-time firmware correlators is that they
produce correlation functions in real time, and avoid the build-up and
storage of huge amounts of photon-count data (e.g., just one of the
present correlators, using its 8 input channels, each running at $50\,MHz$
during one 8-hour observing night would process more than $10\,TB$ of
photon-count data).   Their limitation, of course, is exactly analogous:
if something needs to be checked afterwards, the full set of original data
is no longer available, and alternative signal processing cannot be
applied.

Besides the correlator, another piece of electronics is required for
real-time intensity interferometry, namely one to implement the
continuously variable time delay from the target star to each telescope,
as the star moves across the sky.  Since -- in contrast to the historic
Narrabri interferometer whose telescopes moved on railroad tracks -- the
locations of Cherenkov telescopes are fixed, there is a need to keep
measurements of the target constant within some nanosecond or so, relative
to its wavefront. At Lund Observatory, an experimental hardware unit 
(ÒQVANTOS precision delayÓ) was designed by engineer Bo Nilsson, and verified in the
laboratory.  This unit enables a continuously variable and programmable
delay up to a few $\mu s$ (corresponding to differential light travel distances
of half a km) as applied to a stream of electronic pulses.  The pulses are
read into a large buffer memory and almost simultaneously read out but
then with a slightly different readout frequency, slightly stretching or
squeezing the electronic pulsewidths to create the required delay.
If such a delay unit is not used, the maximum correlation signal will
appear not in the channel for zero time delay between any pair of
telescopes, but rather at a delay equal to the light-time difference
between telescopes along the line of sight towards the source.  In
principle, this would be possible to handle already with existing digital
correlators since these can be programmed to measure the correlation at
full time resolution also at time coordinates away from zero.  However,
since these delays continuously change as the star moves overhead, such an
arrangement would require frequent readouts and might prove not practical
for more routine observations.

\subsection{Correlation using continuous digitization \&  software processing}
Recent advances in commercial high speed data acquisition has enabled the ability to perform the correlation between the telescope intensity
 signals entirely in software.  Using a National Instruments PXIe-5122  high speed digitizer ($100\,MHz$ sampling, 12 bit resolution, 2 channels),
we can continuously digitize the electronic signal stream from each photomultiplier tube. The high speed backplane of the PXIe mainframe
(4 giga-samples per second -- $Gs/sec$) allows the entire data stream to be continuously recorded for several hours onto a high speed RAID disk  ($600\,Mb/sec$,
 NI 6282 Controller +  NI 8264 RAID disk) .  The data streams from multiple telescopes can then be cross-correlated using a software
 correlation algorithm, and digital filtering algorithms  can be employed to implement narrow-band notch filters to eliminate interference
noise from narrow band  sources such as cell-phones,  motors, computer clocks, etc. This type of systems allows measurement
 of  two-telescope intensity correlations as well as higher order correlations (3 telescopes and higher) which may prove  useful for phase recovery.

We are currently testing a two channel continuous digitization stream system built with the above commercial components.
The system has demonstrated the ability to stream data continuously to disk for 4 hours at $100\,MHz$ sampling rate (2 channels,
 12 bit resolution) without loss of any data.  Preliminary FFT studies of the data stream indicate very flat spectral response
and only a few very narrow band man-made noise sources. A 2-telescope software correlation algorithm has been developed,
and we are currently fine tuning the algorithm to develop optimal sensitivity.

During the summer of 2010, we expect to field test the continuous digization system at the StarBase telescopes.  National Instruments
 is also  developing a higher speed continuous streaming digitizer ($500\,MHz$ 8-bit resolution, and $1\,GHz$, 8-bit resolution)
which is scheduled to be available during 3Q 2010.   The imminent availability of these higher speed
commercial digitizers, in conjunction with a substantial reduction in the cost of $10\,Gs/sec$  data transmission
using 9/125 single-mode optical fiber may allow a distributed array to stream data from every telescope
 in an array to a central station for continuous software correlation between every telescope pair in quasi-real time.

The continuous digitization/software correlation approach has  strong advantages, including short development time, excellent flexibility and
ease of modification,  the ability to examine FFT  characteristics of each photomultiplier tube, and the ability to look at higher
order intensity correlations.  These extraordinary advantages  are offset  by the computational  difficulties encountered in handling large
(multi $TB$) data sets. The net result is a very powerful technique which is already strongly competitive with traditional analog and chip-based  correlators.

\section{Outlook}
\label{outlook}
During the summer of 2010, the camera electronics, front end and slow control should be in place to be tested on the StarBase telescopes so signals will be available for correlation studies using the various techniques available. 

Using Equation \ref{snratio} with conservative parameters for the StarBase telescopes ($A=6\,m^2$, $\alpha=0.2$ and $\Delta f=100\,MHz$), the 5 standard deviation measurement of a degree of coherence $|\gamma(d)|^2=0.5$ will require an observation time $T\approx 10\,min \times 2.5^{2V}$ where $V$ represents the visual magnitude and where we have made the crude approximation $n=5\times 10^{-5} \times 2.5^{-V} m^{-2}\cdot s^{-1} \cdot Hz^{-1}$. This corresponds to one hour for $V=1$ and 6.5 hours for V=2 and when considering the measurement of $|\gamma(d)|^2\approx 1$, these observation times should be divided by four. 

The first objective will be the detection of the degree of coherence for an unresolved object ($|\gamma(d)|^2\approx 1$). The distance between the two telescopes being $23\,m$ (smaller baselines can be obtained during observations to the East and to the West due to the projection effect), at $\lambda=400\,nm$ the stars have to be smaller than typically $\sim 3 \, mas$ in diameter. An essentially unresolved star suitable for calibration should be less than $\sim 1\,mas$ in diameter. Good candidates for this are in increasing order of magnitude $\alpha\,Leo$, $\gamma\,Ori$, $\beta\,Tau$ or even $\eta\,UMa$ which, with a magnitude of 1.76 should be observable as a unresolved object for calibration within 50 minutes. Alternatively, it will be possible to measure any star as an unresolved object by correlating the signals from two channels mounted on the same telescope by means of the camera beam splitter. These observations should allow to establish methods for adjusting the signal time delays optimally and also to identify the most effective correlator. A next phase will be dedicated to the measurement of a few bright stars in order to further demonstrate the technique. This second phase will possibly include the observation of coherence modulation resulting from orbital motion in the binary star Spica with a $a=1.5\,mas$ semi major axis and $V=1.0$, or even, possibly Algol ($a=2.18\,mas$, $V=2.1$).

These tests over the coming year or two will be ported to the VERITAS telescopes and will permit to clearly identify all the aspects of a larger scale implementation and converge with confidence toward a practical and effective design of SII instrumentation to be integrated into future IACT arrays such as CTA and AGIS.

\section{Acknowledgement}
This work is supported by  grants SGER \#0808636 from the National Science Foundation. The work at Lund Observatory is supported by the Swedish Research Council and The Royal Physiographic Society  in Lund. The authors are grateful to Linda Nelson and George Sanders for 
hosting the StarBase observatory on the site of the Bonneville Seabase diving resort \cite{seabase} .

\end{document}